\documentstyle[12pt,aaspp4,flushrt]{article}
\makeatletter

\@ifundefined{chapter}{\def\thebibliography#1{\section*{References\@mkboth
  {REFERENCES}{REFERENCES}}\list
  {\relax}{\setlength{\labelsep}{0em}
        \setlength{\itemindent}{-\bibhang}
        \setlength{\itemsep}{0pt}
        \setlength{\parsep}{0pt}
        \setlength{\leftmargin}{\bibhang}}
    \def\newblock{\hskip .11em plus .33em minus .07em}
    \sloppy\clubpenalty4000\widowpenalty4000
    \sfcode`\.=1000\relax}}%
{\def\thebibliography#1{\chapter*{Bibliography\@mkboth
  {BIBLIOGRAPHY}{BIBLIOGRAPHY}}\list
  {\relax}{\setlength{\labelsep}{0em}
        \setlength{\itemindent}{-\bibhang}
        \setlength{\itemsep}{0pt}
        \setlength{\parsep}{0pt}
        \setlength{\leftmargin}{\bibhang}}
    \def\newblock{\hskip .11em plus .33em minus .07em}
    \sloppy\clubpenalty4000\widowpenalty4000
    \sfcode`\.=1000\relax}}

\newlength{\bibhang}
\let\@internalcite\cite
\def\cite{\let\@citeleft(\let\@citeright)%
    \@ifstar{\citeyear}{\citefull}}
\def\acite{\let\@citeleft\relax\let\@citeright\relax%
    \@ifstar{\citeyear}{\acitefull}}
\def\citenp{\let\@citeleft\relax\let\@citeright\relax
    \@ifstar{\citeyear}{\citefull}}
\def\citefull{\def\astroncite##1##2{##1~##2}\@internalcite}
\def\citeyear{\def\astroncite##1##2{##2}\@internalcite}
\def\acitefull{\def\astroncite##1##2{##1~(##2)}\@internalcite}

\def\@citex[#1]#2{\if@filesw\immediate\write\@auxout{\string\citation{#2}}\fi
  \def\@citea{}\@cite{\@for\@citeb:=#2\do
    {\@citea\def\@citea{; }\@ifundefined
       {b@\@citeb}{{\bf ?}\@warning
       {Citation `\@citeb' on page \thepage \space undefined}}%
{\csname b@\@citeb\endcsname}}}{#1}}

\def\@cite#1#2{\@citeleft#1\if@tempswa , #2\fi\@citeright}
\def\@biblabel#1{}

\makeatother

\def\refsponse{{\em RefSponse}}
\received{}
\revised{}
\slugcomment{}

\begin{document}

\title{\refsponse : A Literature Evaluation System for the Professional
    Astrophysics Community}
\author{
Robert E.\ Rutledge\altaffilmark{1} 
\altaffiltext{1}{
Department of Physics, McGill University, 3600 rue University,
    Montreal, Quebec, H3A 2T8, Canada
}
}

\begin{abstract}

We describe an implementation of a semi-automated review system for the
astrophysics literature.  Registered users identify names under which
they publish, and provide scores for individual papers of their
choosing.  Scores are held confidentially, and combined in a weighted
average grade for each paper.  The grade is divided among the
co-authors as assigned credit.  The credit accumulated by each user
(their ``mass'') provides the weight by which their score is averaged
into papers' grades.  Thus, papers' grades and users' masses are
mutually dependent and evolve in time as scores are added.  Likewise,
a user's influence on the grade of a paper is determined from the
perceived original scientific contribution of all the user's previous
papers.  The implementation, called {\em RefSponse} -- currently
hosted at http://bororo.physics.mcgill.ca -- includes papers in
astro-ph, the ApJ, AJ, A\&A, MNRAS, PASP, PASJ, New Astronomy, Nature,
ARA\&A, Phys.\ Rev.\ Letters, Phys.\ Rev.\ D.\ and Acta Astronomica
from 1965 to the present, making extensive use of the NASA/ADS
abstract server.  We describe some of the possible utilities of this
system in enabling progress in the field.
\end{abstract}


\section{Introduction}

Electronic distribution of the academic literature has opened a
disjoint between the time when papers are introduced to a public
readership, and their appearance in peer-reviewed journals -- the
specific event which historically has marked the paper as providing a
minimal original academic contribution.  This disjoint, its
implications for progress in the field, and some possible solutions,
have been been discussed elsewhere (Ginsparg 2002).

The web's ability to speed and broaden communications and its
application to the scientific literature -- most famously via arXiv --
has been widely exploited to the benefit of the scientific community.
In addition, the web provides means to aggregate quantitative
evaluations.  However, 10 years after the introduction of the web, the
astrophysics community has not exploited evaluation aggregation in
application to its literature.

The applications and utilities of aggregating opinions of the
distributed astrophysics literature are unexplored.  Here, we
introduce an implemented method of aggregating opinions of the
astrophysics literature.  We discuss some of the useful functions it
may serve, and draw distinctions between these functions and those of
detailed journal peer review.  We close by pointing out that an
aggregated numerical grade cannot replace specific consideration of
the merits and flaws of papers in the literature, which is the only
basis for progress; but the existence of an aggregate grade will
permit individuals to identify papers about which there exists wide
misconception (as evidenced by the grade), and to work to correct
this, enabling progress in the field.

\section{Operational Principles}
\label{sec:op}

The basic operational principles are: 

\begin{itemize}
\item The primary goal of all operations is to encourage the writing
of outstanding papers in astrophysics. 
\item The primary purpose of usage is to permit individual users to function
as a community, scoring individual papers in proportion to their
original scientific contribution, and having these scores
aggregated into a grade which represents the users' jointly
held opinion. 
\item The weight of an individual user's score should be proportional
  to the sum-total of that user's original scientific contributions. 
\item Users are capable of self-identifying papers which they are
  competent to score within the precision of the 0-20 scoring
  system.  Users are also capable of relying on the sole criterion of
  original scientific contribution when scoring papers.  
\item Individual user's scores are confidential (that is, it is not possible to
determine the score given to any individual paper by any individual
user). 
\end{itemize}

\section{Operational Mechanics}
\label{sec:mech}

In this section, we describe basic mechanics of how the service
operates.  The main operations -- with the exception of reading and
evaluating the papers --- have been automated: 

\begin{enumerate}
\item Identify the journals and collections of papers which contain
the articles to  be included for evaluation; 
\item uniquely identify each article which appears
in these journals (in the \refsponse\ implementation, we use the ADS bibcode);
\item identify all co-authors on all articles; 
\item assign the co-author's identities to system users; 
\item permit users to numerically score articles; 
\item employ quantitative criteria (``certification'') to determine which users shall have their
scores used in calculating papers' grades; 
\item combine scores, weighted by the users' ``mass'', into a grade
  for each paper; 
\item partition the grade of each article to each
co-author of the article (and, in turn, the user assigned the
co-author's identity), and sum the partitioned grades for each
user into the value known as the user's ``mass''.  
\item iteratively solve for the mutually dependent paper grades and
users' ``masses.''

\end{enumerate}

This is similar to the group-moderated message boards in widespread
use on the Internet for a decade.  The main difference here is that
\refsponse\ is designed for specific use with the astrophysics
literature.

We rely for our references database on the publicly held information
from the NASA/ADS Abstracts database.  We ingest information about
individual papers (see \S~\ref{sec:journals}), including titles,
author lists and the ADS uniform bibliography code (bibcode).  These
are databased locally and updated daily.

{\em Users} register at the website for a username+password protected
account.  Registration -- subject to review by the operator -- is at
present open.  At some time in the future, the website will require
single-use registration certificates. These will be distributed to the
user base, for passing along to interested colleagues.  This procedure
will permit initial registrations to proceed quickly, while insuring a
long-term orderly process for expanding the user base. 

On registration, users identify the complete list of names they use to
publish papers, which will identify the papers on which they are
co-author.  

Users may then search for papers in the literature which they
self-identify as competent to review to the precision offered by the
grading scale.  Users may submit a {\em score}, value 0-20, in strict
proportion to their opinion of the original scientific contribution of
the paper.  

Once per day, the scores on each paper are aggregated into a weighted
average, to produce a single numerical {\em grade} for that paper.
Simultaneously, the {\em grades} on every paper co-authored by each
user are aggregated -- roughly speaking, added together -- to
calculate the user's {\em mass}.  Thus, the user's mass is higher
when the user has co-authored more, high quality papers.

For purposes of transparency, after there are $>$100 users registered
to the site, registered users will have access to the list of users,
including only names, institutions, and the ``published under'' names.
This will permit independent authentication of user identities. 

\subsection{Scoring}

Any registered user may score a paper.  Only the scores of users who
have been Certified (see \S~\ref{sec:certification}) are included in
calculating papers' grades.  The score range of 0-20 was chosen as a
compromise between wide dynamic range and reasonable (integral)
resolution in scoring.

To address the concern of score inflation (i.e., a user scoring every
paper a 20), we assume the following regarding the intrinsic
distribution of scores.  We assume that two factors are involved of
producing a paper of score $S$: time and skill.  The distribution of
scores due to time investment, skill being equal, means there should
be 20$\times$ as many papers with ``1'' as there are ``20''
($N\propto1/S$).  The distribution of scores for skill, time
investment being equal, is a similar factor ($N\propto1/S$).  And,
finally, that skill and time are directly proportional -- that
doubling the amount of applied skill and halving the applied time to a
paper results in a work deserving an equal grade.  We thus assume that
the intrinsic distribution of scores should be $N(S)\propto1/S^2$.

We enforce this scoring distribution through the minimal requirement
that a user's non-zero scores average remain below 2.25.  Doing so
also prompts users to re-visit scores on older papers, to determine if
the perception of their relative contribution is sustained. 

\subsection{User Certification}
\label{sec:certification}

A user's scores will be folded into the paper grade only if the user
meets certain minimum requirements (``certification''): 

\begin{itemize}
\item scores on $\geq$3 papers on which the user is a co-author, placed
by $\geq$3 different users. 
\item The user must have scored $>$10 papers within the previous
calendar year.  
\item Certification is performed during each grading. 
\item Users are de-certified if they do not meet the above
criteria. 
\end{itemize}

Using this quantitative criterion for including a user's score in a
paper grade, as well as using the relative weighting of a user's mass
in determining a paper's grade, permits automation of determining the
paper's grade. 

\subsection{Calculation of Paper Grades and User Masses}

The calculation of paper grades ($G_i$) and user masses ($M_j$) is
non-analytic and so solved iteratively, relying on user scores for
each paper ($S_{i,j}$).

The paper grade is a weighted average of the users' scores:

\begin{equation}
G_i = \frac{ \sum_j  M_j S_{i,j}}{ \sum_j M_j}
\end{equation}

The weight for this average is the user's mass ($M_j$).  The paper
grade takes a value between the extremes of assigned scores.  Each
user's mass is found:

\begin{equation}
M_j = \sum_i N_{n_j,i} G_i
\end{equation}

\noindent where $N_{n_j,i}$ is the fraction of the grade $G_i$
attributed to user $j$ when they are author number $n_j$ on paper
$i$.\footnote{First author is $n_j=1$, second author is $n_j=2$, and so
on.}.  Note $\sum_j N_{n_j}=1$. 

As a user's mass increases (as it will, in time, as they co-author
more papers), their relative influence on a paper's grade
also increases.  Thus, the grades of papers will evolve in response to
evolution in users' masses, even while the scores themselves do not
change.  The operating assumption is that users who have written many,
highly regarded papers should have proportional influence in
determining the grade of others' papers.

\subsubsection{The Normalization $N_{n_j}$}

The normalization $N_{n_j}$ determines how the credit (represented by
the grade) of a paper is divided among its co-authors.  This is an
important function for operations, worthy of some discussion and
consideration.  The criteria we use to evaluate methods of calculating
$N_{n_j}$ are: (1) that it reflect reasonable practices of
distributing credit for individual papers; and (2) that it need not
duplicate existing practices for attributing credit to individuals;
(3) automated implementation be practical.

We discuss four different approaches, although there are certainly others:

\begin{itemize}
\item {\bf Equal Credit per Author (ECA)}: For $n_{\rm
authors}$ authors on the author-list, $N_{n_j}$=1/$n_{\rm authors}$,
the credit is divided equally among all authors. 
\item {\bf First author leads, others divide (FAL)}:  $N_{n_j=1}= n_{\rm authors}/(2n_{\rm authors} -1)$ and
$N_{n_j>1}=1/(2N_{\rm authors} -1)$.  
\item {\bf Author Determines (AUT)}: The first author selects from among
these options, or provides an arbitrary distribution of $N_{n_j}$.
\item {\bf A deprecated grading per author (DEP)}.  Setting
$\chi=\ln(N_{\rm authors})/\ln(2)$, then: 
\begin{eqnarray}
N_{n_j}  = & \frac{1}{2^{n_j}} & n_j \leq \chi\ \\
	 = & \frac{1}{2^{{\rm int}(\chi)}(N_{\rm authors}-{\rm int}(\chi))} & n_j>\chi
\end{eqnarray}

\noindent where ${\rm int}(\chi)$ is the truncation of $\chi$ to an integer. 

\end{itemize}

While ECA may seem equitable, some authors put more work into papers
than others. Most typically, the authors' names appearing to the front
of the graded paper (first author, second author) lead the effort.
The ECA method also duplicates the practice of attributing credit by
counting the number of papers co-authored.

The FAL method recognizes the first author, and treats all remaining
authors equally.  In papers with many authors, it fails to offer
significant credit to any but the first author.  

The AUT method most probably reflects the fairest method, since most
likely the first-author will be most able to determine the fair
distribution of each paper's credit.  However, it would require the
cooperation of all first authors of all published papers, which is not
practical.

DEP has the advantage of recognizing the first and early authors of
the paper, reflecting a wide practice of ordering authors by their
contributions.  DEP fails when authors are ordered by other criteria
-- for example, alphabetization -- which would require AUT. 

Of the four methods, by the criteria above, we find the most
preferable method for determining $N_{n_j}$ is DEP, which we adopt for
\refsponse.  However, the method is not exclusive -- it's possible to
introduce $>$1 method to run in parallel. 

\subsection{Included Journals}
\label{sec:journals}
Journals which are included in this service are: 

\begin{itemize}
\item astro-ph (since 2003). 
\item ApJ, Letters, Supplements
\item A\&A, Letters, Supplements
\item AJ
\item Annual Review of Astronomy and Astrophysics
\item MNRAS
\item Nature
\item New Astronomy
\item PASP
\item PASJ 
\item Acta Astronomica
\item Physical Review, D
\item Physical Review, Letters
\end{itemize}

We include articles from 1965 to the present, except as noted.  Since
papers which appear on astro-ph often later appear in other journals,
those which are bibliographed as such by the ADS are updated within
\refsponse.

\subsection{User Experience}

A paper's grade reflects the community aggregate opinion of the paper,
and should not be thought of as the ``final word'' on its original
scientific contribution; opinion evolves in response to new
information.  If the grade of a paper is an ``8'', while a user is
aware of a fatal flaw meriting a ``0'', then the user has identified a
paper about which the flaw may not be widely known. But perhaps the
paper was scored by users with little mass (the misconception is not
so wide).  If so, merely adding their own score may drop the grade to
the more appropriate level.  

On the other hand, the flaw may not be widely known, and the user's
mass may be small compared with the sum of all other scorers, in which
case the user's score will have little effect on the paper's grade.
In that case, the user has identified a subject on which wide
misconception exists, which should be addressed by a detailed paper.

\section{Discussion}

\subsection{Comparison Between Peer Review and \refsponse}

\refsponse\ is operationally different from peer review, and should
not be expected to perform identical functions.   Major differences
between \refsponse\ and peer review include:

\begin{enumerate}
\item Peer-review is managed as a qualitative evaluation of the
original scientific contribution presented by an article, providing a
binary outcome (either the paper is included in the journal or
collection. or it is not).  In contrast, \refsponse\ is quantitative,
and provides a graded scale of outcomes which permits relative
comparisons between articles.

\item Peer-review requires a skilled and trusted editor
whose broad knowledge permits identification of individual reviewers
with the scientific capability to evaluate paper quality and with an
absence of motivation to evaluate the paper on any criterion but its
original scientific content.  \refsponse\ uses a minimal qualification
for evaluators (the ``certification'' process), assumes that users
will self-identify their capability to evaluate specific papers, and
will absent themselves any judgment criteria other than original
scientific contribution.

\item Peer-review often places responsibility of evaluation in an
individual, which represents a ``single point of failure'' for that
process.  \refsponse\ diffuses responsibility among many individuals,
giving greater influence to those who have greater
recognition of their original scientific contributions.  

\item Peer-review can address issues other than paper quality, such as
academic integrity.  No similar capability exists for \refsponse.

\end{enumerate}

\refsponse\ can provide the basis for some useful services.  For
example, if \refsponse\ operates successfully it will be possible to:

\begin{enumerate}
\item Determine a rough evaluation of papers which have not
  entered (or will not enter) the peer-reviewed journals. 
\item Obtain a grade-sorted list of all papers of a specific year or a
  specific journal, and immediately know where the important work took
  place in a particular year.
\item Search for highly-graded papers by subject area. 
\item Produce a list of high-graded papers scored by users who scored
  the same papers as you -- but which you have not scored.  This would
  call your  attention to papers which you perhaps under-appreciate,
  or which colleagues over-appreciate.  
\item Determine the perceived aggregate scientific contribution of
  specific journals, and so determine their individual
  cost-effectiveness.
\item For novices in the field, provide a first means of filtering
  papers for quality, to focus their reading toward those with
  perceived contributions. 
\item Provide a simple quantitative measure of the quality of one's
  own work, and monitor its progress. 
\item Have a scientific ``bake off''.  In an
  ongoing public scientific debate, two (or more) papers can be
  compared, and the community evaluation of each determined. 
\end{enumerate}

In closing, it should be pointed out that this method cannot replace
reasoned discussion of the merits and flaws of papers in the
literature; such discussion is the only basis for progress.

However, aggregating the community's opinion as a grade can help
provide that basis for progress.  In the example of the scientific
``bake off'', simply comparing the grades of the two papers will not
teach one anything about their content.  However, if one also studies
the two papers and forms an opinion of their relative merit opposite
that of the community's, then a misconception clearly exists -- either
one's own, or the community's.  Identifying and eradicating the source
of that misconception is exactly what is meant by ``progress''.

\end{document}